\documentstyle[psfig]{aipproc}

\def\la{\mathrel{\hbox{\rlap{\hbox{\lower4pt\hbox{$\sim$}}}\hbox{$<$}}}}
\def\ga{\mathrel{\hbox{\rlap{\hbox{\lower4pt\hbox{$\sim$}}}\hbox{$>$}}}}

\begin{document}
\title{The Detectability of Gamma-Ray Bursts and Their 
Afterglows at Very High Redshifts}

\author{Donald Q. Lamb and Daniel E. Reichart} 
\address{Department of Astronomy \& Astrophysics, University of Chicago,
\\ 5640 South Ellis Avenue, Chicago, IL 60637}

\maketitle

\begin{abstract}
There is increasingly strong evidence that gamma-ray bursts (GRBs) are
associated with star-forming galaxies, and occur near or in the
star-forming regions of these galaxies.  These associations provide
indirect evidence that at least the long GRBs detected by BeppoSAX are
a result of the collapse of massive stars.  The recent evidence that
the light curves and the spectra of the afterglows of GRB 970228 and
GRB 980326 appear to contain a supernova component, in addition to a
relativistic shock wave component, provide more direct clues that this
is the case.  Here we establish that GRBs and their afterglows are both
detectable out to very high redshifts ($z \gtrsim 5$).
%Consequently, if many GRBs are indeed produced  by the collapse of
%massive stars, GRBs and their afterglows provide a powerful probe of
%the very high redshift universe.  
\end{abstract}

\section*{Introduction} 

We first show that the GRBs with well-established redshifts could have been
detected out to very high redshifts (VHRs).  Then, we show that their
soft X-ray, optical, and infrared afterglows could also have been
detected out to these redshifts.

\section*{Detectability of GRBs} 

We first show that GRBs are detectable out to very high redshifts.  The
peak photon number luminosity is \begin{equation} L_P =
\int_{\nu_l}^{\nu_u}\frac{dL_P}{d\nu}d\nu \; , \end{equation} where
$\nu_l < \nu < \nu_u$ is the band of observation.  Typically,  for
BATSE, $\nu_l = 50$ keV and $\nu_u = 300$ keV.  The corresponding peak
photon number flux $P$ is  \begin{equation} P =
\int_{\nu_l}^{\nu_u}\frac{dP}{d\nu}d\nu \; . \end{equation} Assuming
that GRBs have a photon number spectrum of the form $dL_P/d\nu \propto
\nu^{-\alpha}$ and that $L_P$ is independent of z, the observed peak
photon number flux $P$ for a burst occurring at a redshift $z$ is given
by \begin{equation} P = \frac{L_P}{4\pi D^2(z)(1+z)^{\alpha}} \; ,
\end{equation} where $D(z)$ is the comoving distance to the GRB. 
Taking $\alpha = 1$, which is  typical of GRBs [1], Equation (3) coincidentally reduces to the form that one gets
when $P$ and $L_P$ are
bolometric quantities.

Using these expressions, we have calculated the limiting redshifts
detectable by BATSE and HETE-2, and by {\it Swift}, for the seven GRBs
with well-established redshifts and published peak photon number
fluxes.  In doing so, we have used the peak photon number fluxes given
in Table 1 of [2], taken a detection threshold of 0.2 ph s$^{-1}$ for BATSE
[3] and HETE-2 [4]  and 0.04 ph s$^{-1}$ for
{\it Swift} [5], and set $H_0 =  65$ km s$^{-1}$ Mpc$^{-1}$,
$\Omega_m = 0.3$, and $\Omega_{\Lambda} = 0.7$ (other cosmologies give
similar results).

\begin{figure}
\begin{minipage}[t]{2.75truein}
\mbox{}\\
\psfig{file=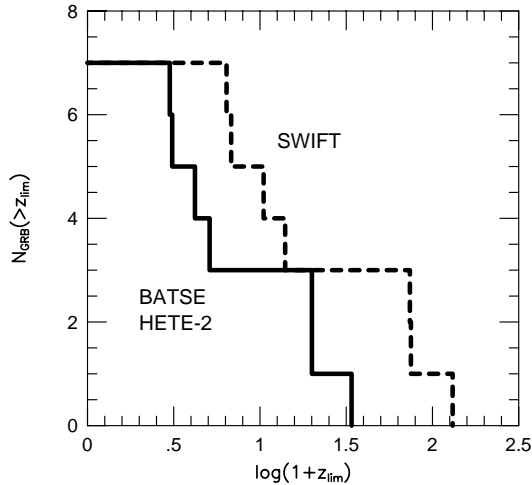,width=2.75truein,clip=}
\end{minipage}
\hfill
\begin{minipage}[t]{2.75truein}
\mbox{}\\
\caption{Cumulative distributions of the limiting redshifts at which
the seven GRBs with well-determined redshifts and published peak photon
number fluxes would be detectable by BATSE and HETE-2, and by {\it
Swift}.}
\end{minipage}
\end{figure}

Figure 1 displays the results.  This figure shows that BATSE and
HETE-2  would be able to detect four of these GRBs (GRBs 970228,
970508, 980613, and 980703) out to redshifts $2 \lesssim z \lesssim 4$,
and three (GRBs 971214, 990123, and 990510) out to redshifts of $20
\lesssim z \lesssim 30$. {\it Swift} would be able to detect the former
four out to redshifts of $5 \lesssim z \lesssim 15$, and the latter
three out to redshifts in excess of $z \approx 70$, although it is
unlikely that GRBs occur at such extreme redshifts (see \S 3 below). 
Consequently, if GRBs occur at VHRs, BATSE has probably already
detected them, and future missions should detect them as well.

\section*{Detectability of GRB Afterglows} 

The soft X-ray, optical and infrared afterglows of GRBs are also
detectable out to VHRs.  The effects of distance and redshift tend to
reduce the spectral flux in GRB afterglows in a given frequency band,
but time dilation tends to increase it at a fixed time of observation
after the GRB, since afterglow intensities tend to decrease with time. 
These effects combine to produce little or no decrease in the spectral
energy flux $F_{\nu}$ of GRB afterglows in a given frequency band and
at a fixed time of observation after the GRB with increasing redshift:
\begin{equation}
F_{\nu}(\nu,t) = \frac{L_{\nu}(\nu,t)}{4\pi D^2(z) (1+z)^{1-a+b}},
\end{equation}
where $L_\nu \propto \nu^at^b$ is the intrinsic spectral luminosity of
the GRB afterglow, which we assume applies even at early times, and
$D(z)$ is again the comoving distance to the burst.   Many afterglows
fade like $b \approx -4/3$, which implies that $F_{\nu}(\nu,t) \propto
D(z)^{-2} (1+z)^{-5/9}$ in the simplest afterglow model where $a =
2b/3$ [6].  In addition,
$D(z)$ increases very slowly with redshift at redshifts greater than a
few.  Consequently, there is little or no decrease in the spectral flux
of GRB afterglows with increasing redshift beyond $z \approx 3$.  

For example, [7] find in the case of GRB 980519 that
$a = -1.05\pm0.10$ and $b = -2.05\pm0.04$ so that $1-a+b = 0.00 \pm
0.11$, which implies no decrease in the spectral flux with increasing
redshift, except for the effect of $D(z)$.  In the simplest afterglow
model where $a = 2b/3$, if the afterglow declines more rapidly than $b
\approx 1.7$, the spectral flux actually {\it increases} as one moves
the burst to higher redshifts! 

As another example, we calculate the best-fit spectral flux
distribution of the early afterglow of GRB 970228 from [8], as observed one day after the burst, transformed to various
redshifts.  The transformation involves (1) dimming the
afterglow,\footnote{Again, we have set $\Omega_m = 0.3$ and
$\Omega_{\Lambda} = 0.7$; other cosmologies yield similar results.} (2)
redshifting its spectrum, (3) time dilating its light curve, and (4)
extinguishing the spectrum using a model of the Ly$\alpha$ forest.  
For the model of the Ly$\alpha$ forest, we have adopted the best-fit
flux deficit distribution to Sample 4 of [9] from [10].  At redshifts in excess of $z = 4.4$, this model is an
extrapolation, but it is consistent with the results of theoretical
calculations of the redshift evolution of Ly$\alpha$ absorbers [11].  Finally, we have convolved
the transformed spectra with a top hat smearing function of width
$\Delta \nu = 0.2\nu$.  This models these spectra as they would be
sampled photometrically, as opposed to spectroscopically; i.e., this
transforms the model spectra into model spectral flux distributions.

Figure 2 shows the resulting K-band light curves.  For a fixed
band and time of observation, steps (1) and (2) above dim the afterglow
and step (3) brightens it, as discussed above.  Figure 2 shows that in
the case of the early afterglow of GRB 970228, as in the case of GRB
980519, at redshifts greater than a few the three effects nearly cancel
one another out.  Thus the afterglow of a GRB occurring at a redshift
slightly in excess of $z = 10$ would be detectable at K $\approx 16.2$
mag one hour after the burst, and at K $\approx 21.6$ mag one day after
the burst, if its afterglow were similar to that of GRB 970228 (a
relatively faint afterglow).  

Figure 3 shows the resulting spectral flux distribution.  The spectral
flux distribution of the afterglow is cut off by the Ly$\alpha$ forest
at progressively lower frequencies as one moves out in redshift.  Thus 
high redshift ($1 \lesssim z \lesssim 5$) afterglows are characterized
by an optical ``dropout'' [12], and very high redshift ($z
\gtrsim 5$) afterglows by an infrared ``dropout.''

%We also show in Figure 3 the effect of a moderate ($A_V = 1/3$), fixed
%amount of extinction at the redshift of the GRB.  However, the amount
%of extinction is likely to be very small at large redshifts because of
%the rapid decrease in metallicity beyond $z = 3$.

In conclusion, if GRBs occur at very high redshifts, both they and
their afterglows would be detectable.

\begin{figure}
\begin{minipage}[t]{2.75truein}
\mbox{}\\
\psfig{file=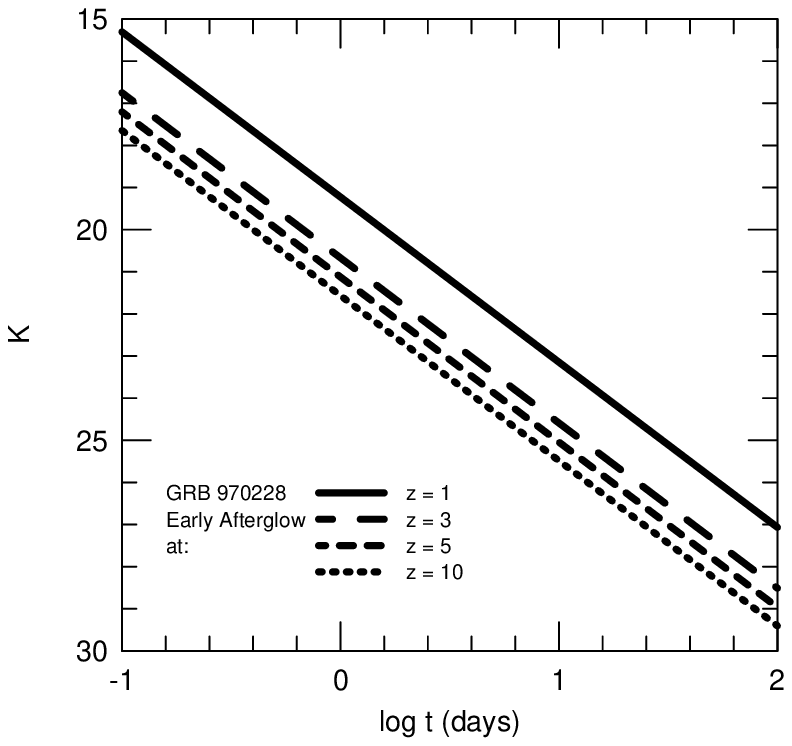,width=2.75truein,clip=}
\caption{The best-fit light curve of the early
afterglow of GRB 970228 from Reichart (1999), transformed to various
redshifts.}
\end{minipage}
\hfill
\begin{minipage}[t]{2.75truein}
\mbox{}\\
\psfig{file=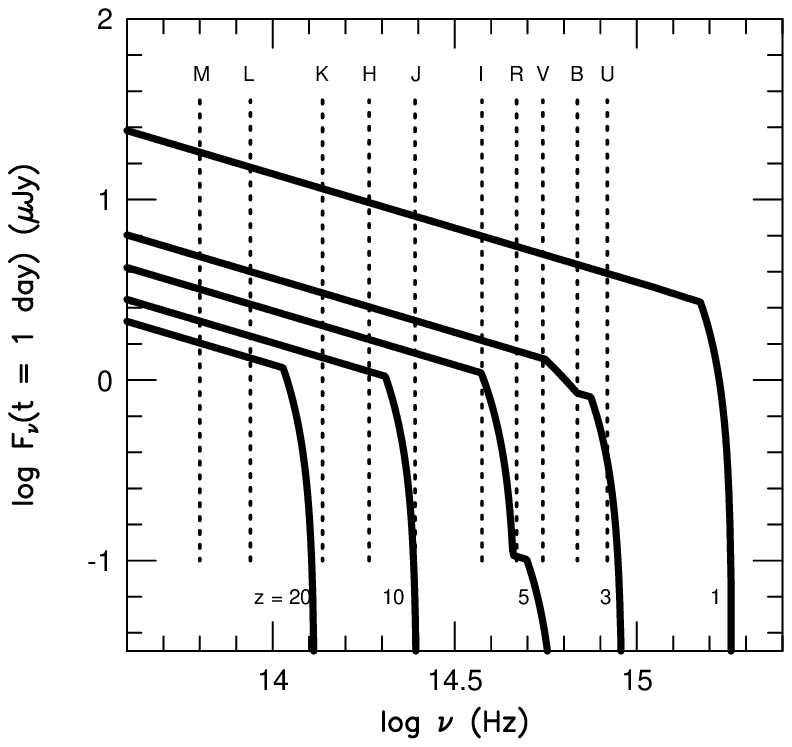,width=2.75truein,clip=}
\caption{The best-fit spectral flux distribution of the
early afterglow of GRB 970228 from Reichart (1999), as observed one
day after the burst, after transforming it to various redshifts, and
extinguishing it with a model of the Ly$\alpha$ forest.}
\end{minipage}
\end{figure}

\end{document}